\documentclass[english,aps,superscriptaddress,preprintnumbers]{revtex4}
\usepackage[T1]{fontenc}
\usepackage[latin9]{inputenc}
\usepackage{babel}
\usepackage{verbatim}
\usepackage{amsmath}
\usepackage{amssymb}
\usepackage{graphicx}
\usepackage{color}
\usepackage[unicode=true,pdfusetitle,
 bookmarks=true,bookmarksnumbered=false,bookmarksopen=false,
 breaklinks=false,pdfborder={0 0 1},backref=false,colorlinks=false]
 {hyperref}

\makeatletter
\@ifundefined{textcolor}{}
{%
 \definecolor{BLACK}{gray}{0}
 \definecolor{WHITE}{gray}{1}
 \definecolor{RED}{rgb}{1,0,0}
 \definecolor{GREEN}{rgb}{0,1,0}
 \definecolor{BLUE}{rgb}{0,0,1}
 \definecolor{CYAN}{cmyk}{1,0,0,0}
 \definecolor{MAGENTA}{cmyk}{0,1,0,0}
 \definecolor{YELLOW}{cmyk}{0,0,1,0}
}

\usepackage{babel}

\makeatother

\begin{document}

\title{Wigner function and pair production in parallel electric and magnetic fields}

\author{Xin-li Sheng}

\affiliation{Interdisciplinary Center for Theoretical Study and Department of
Modern Physics, University of Science and Technology of China, Hefei,
Anhui 230026, China}
\affiliation{Institute for Theoretical Physics, Goethe University, Max-von-Laue-Str.\ 1,
D-60438 Frankfurt am Main, Germany}

\author{Ren-hong Fang}

\affiliation{Key Laboratory of Quark and Lepton Physics (MOE) and Institute of
Particle Physics, Central China Normal University, Wuhan, Hubei 430079,
China}

\author{Qun Wang}

\affiliation{Interdisciplinary Center for Theoretical Study and Department of
Modern Physics, University of Science and Technology of China, Hefei,
Anhui 230026, China}

\author{Dirk H.\ Rischke}

\affiliation{Institute for Theoretical Physics, Goethe University, Max-von-Laue-Str.\ 1,
D-60438 Frankfurt am Main, Germany}

\affiliation{Interdisciplinary Center for Theoretical Study and Department of
Modern Physics, University of Science and Technology of China, Hefei,
Anhui 230026, China}
\begin{abstract}
We derive analytical formulas for the equal-time Wigner function
in an electromagnetic field of arbitrary strength. While the magnetic field
is assumed to be constant, the electric field is assumed to be
space-independent and oriented parallel to the magnetic field. The Wigner
function is first decomposed in terms of the so-called Dirac-Heisenberg-Wigner
(DHW) functions and then the
transverse-momentum dependence is separated using a new set of basis functions
which depend on the quantum number $n$ of the Landau levels. Equations for the coefficients
are derived and then solved for the case of a constant electric field. The
pair-production rate for each Landau level is calculated. In the case
of finite temperature and chemical potential, the pair-production
rate is suppressed by Pauli's exclusion principle.
\end{abstract}

\preprint{\hfill {\small {ICTS-USTC-18-20}}}

\maketitle

\section{Introduction}

Quantum electrodynamics (QED) of strong electromagnetic (EM) fields has been studied for a very
long time \cite{Greiner:1985ce}. In the initial stage of non-central heavy-ion collisions, the electromagnetic
field can be as large as $9.8\times10^{22}\, \mathrm{V}/\mathrm{m}$
at the Relativistic Heavy Ion Collider (RHIC) \cite{Kharzeev:2007jp} and even larger at
the Large Hadron Collider (LHC). Such extremely
strong fields are generated by the fast-moving nuclei but will rapidly
fall off with time \cite{Rafelski:1975rf}. The medium was estimated to
extend the lifetime of the fields and enhance the possibility of detecting
the influence of strong EM fields \cite{Tuchin:2014iua,Li:2016tel}.

A strong magnetic field leads to interesting effects related to the chiral anomaly of quantum chromodynamics (QCD).
On the other hand, a strong electric field can lead to decay of the QED
vacuum. When the field strength is near or above the critical strength
$E_{c}=m^{2}c^{3}/q\hbar$ \cite{Sauter:1931zz,Heisenberg:1935qt,Schwinger:1951nm}, where
$m$ is the mass of the particle and $q$ its electric charge,
particle-antiparticle pairs can be created from vacuum.
This process is commonly called Schwinger process in honor of Julian
Schwinger, who derived the pair-production rate in a famous work \cite{Schwinger:1951nm}.
The rate is exponentially suppressed below the critical field strength,
which is about $1.32\times10^{18}\, \mathrm{V}/\mathrm{m}$ for electron-positron
production. Pair production is a nonlinear phenomenon and the corresponding
experiment is important for studying QED beyond the perturbative regime. 
The mechanism can occur in many systems such as in the early Universe,
around neutron stars, and in heavy-ion collisions, while it is
expected to appear in strong-laser experiments like the free-electron X-ray
laser XFEL \cite{Ringwald:2001ib,Ringwald:2001cp} and the extreme-light
infrastructure ELI \cite{Dunne:2008kc}.

Although the Schwinger process has been studied for more than half
a century, calculating the pair production in an arbitrary electromagnetic
field is still a challenging problem. The case of a vanishing magnetic
field $\mathbf{B}(t,\mathbf{x})=0$ and a space-independent electric
field has been exhaustively discussed, where the problem can be translated
into solving the famous Vlasov equation of quantum kinetic theory
\cite{Smolyansky:1997fc,Schmidt:1998vi,Kluger:1998bm}.
It can be analytically solved for a constant electric field $E(t)=E_{0}$
and the Sauter-type field $E(t)=E_{0}\, \mathrm{sech}^{2}(t/\tau)$.
Many theoretical methods are developed to deal with these two cases 
and go beyond these analytical benchmarks, 
such as directly through quantum field theory \cite{Schwinger:1951nm},
WKB methods \cite{Brezin:1970xf,Popov:1971iga,Popov:1973az}, instanton
methods \cite{Affleck:1981bma,Kim:2000un,Dunne:2006ur}, the Wigner-function
method \cite{Hebenstreit:2010vz,Kohlfurst:2015niu,Kohlfurst:2017git}, the 
numerical world-line loop method \cite{Gies:2005ym,Schneider:2018huk},
and holographic methods \cite{Ambjorn:2011wz,Sato:2013pxa,Sato:2013dwa}.
In principle, some methods such as the Wigner-function method 
\cite{Hebenstreit:2010vz,Kohlfurst:2015niu,Kohlfurst:2017git} can be applied 
to very general cases, but one faces a system of non-linear partial differential equations.
However, the field configurations in cosmology or in heavy-ion
collisions are much more complicated than the above mentioned cases. 
One might find an approximate solution by partitioning space-time into small cells
and applying the analytical results for a constant electric field in each cell. This, however,
may generate uncontrollable uncertainties because an instanton study \cite{Dunne:2005sx} showed that
temporal inhomogeneities tend to enhance the pair production while
spatial ones tend to suppress it. Especially in heavy-ion collisions,
where the EM fields vary rapidly in both space and time \cite{Deng:2012pc,McLerran:2013hla,Li:2016tel},
a proper numerical treatment is necessary \cite{Blinne:2013via,Berenyi:2013eia}. 

Nowadays many researchers are focusing on the Schwinger process in
strong-laser experiments \cite{Tajima:1900zz,Bulanov:2004de}. The
critical field strength $E_{c}$ for $e^{+}e^{-}$ pair production
corresponds to an average laser intensity $I_{c}=\frac{1}{2\mu_{0}c}E_{c}^{2}\simeq2.3\times10^{29}\,
\mathrm{W/cm^{2}}$.
Unfortunately, such a large intensity is difficult to generate in an experiment. In the
ELI project \cite{Dunne:2008kc}, the laser pulse can only reach $\sim10^{26}\,\mathrm{W/cm^{2}}$,
which is three magnitudes lower than $I_{c}$. The pair production
in such a case is strongly suppressed by a factor $\exp(-\pi E_{c}/E)\simeq10^{-66}$.
Clearly, the critical intensity $I_{c}$ is not attainable for laser
experiments in the near future. Meanwhile, the electric field in heavy-ion collisions can reach
$eE\sim m_{\pi}^{2}c^3/\hbar \gg eE_c$ at RHIC \cite{Kharzeev:2007jp},
which provides realistic conditions to study pair production. In the recently
discovered Dirac semimetals, massless Dirac fermions can be excited
by an external electromagnetic field and may be experimentally observed
through their transport properties. The production rate remains finite
even if the Dirac fermions are massless \cite{Abramchuk:2016afc},
which is different from the Schwinger process in vacuum.

According to Maxwell's equations, a varying electric field will generate a
magnetic field. Analytical calculations show that a magnetic
field which is parallel to the electric field can increase the pair-production rate
\cite{Nikishov:1969tt,Bunkin:1970iz,Popov:1971iga,Daugherty:1976mg,Dunne:2004nc}.
Recently the enhancement of the pair-production rate due to parallel magnetic fields
has been studied in string theory \cite{Lu:2017tnm,Lu:2018nsc,Jia:2018mlr}.
The pair production rate is modified by the thermal medium \cite{Gies:1999vb,Kim:2008em,Gould:2017fve}.
In this paper, we will reproduce these results via the Wigner-function method.
On the other hand, the pair production in parallel electromagnetic
fields is related to the chiral anomaly \cite{Cai:2016jgr,Guo:2016nnq}
and to pseudoscalar condensation \cite{Cao:2015cka,Fang:2016uds,Lin:2018aon}, 
which can be verified using the results of this paper. We will focus
on these effects in future work.

This paper is organized as follows: In Sec.\ \ref{sec:DHW-functions}
we will briefly introduce the equal-time Wigner function and the Dirac-Heisenberg-Wigner
(DHW) functions. General equations of motion for the DHW functions
are also listed in this section. In Sec.\ \ref{sec:In-electric-fields}
we simplify the equations of motion for the DHW functions in a spatially homogeneous electric
field and give analytical solutions for a constant electric field.
A constant magnetic field, which is parallel to the electric field, is taken into account in
Sec.\ \ref{sec:In-parallel-electric}. The DHW functions reflect
the behavior of the Landau levels. Analytical solutions are derived when
both electric and magnetic fields are constant. In Sec.\ \ref{sec:Pair-production-rate}
we read off the pair-production rate from the DHW functions derived
in Sec.\ \ref{sec:In-parallel-electric}. In Sec.\ \ref{sec:Summary}
we give a summary and provide an outlook to future work. Details about the auxiliary
functions used in this paper and their properties are summarized in
App.\ \ref{sec:Auxiliary-functions}.

We take fermions to have positive unit charge $q=+e$ and the electric
and magnetic fields to point in the $z$-direction. We use the following
notations for four-vectors: $X=(x^{\mu})=(t,\mathbf{r})=(t,\mathbf{x}_{T},z)=(t,x,y,z)$
and $P=(p^{\mu})=(E,\mathbf{p})=(E,\mathbf{p}_{T},p_{z})=(E,p_{x},p_{y},p_{z})$.
We also use the following differential operators $\partial_{t}=\frac{\partial}{\partial t}$,
$\boldsymbol{\nabla}_{\mathbf{x}}=(\frac{\partial}{\partial x},\frac{\partial}{\partial z},\frac{\partial}{\partial z})$
and $\boldsymbol{\nabla}_{\mathbf{p}}=(\partial_{p_{x}},\partial_{p_{y}},\partial_{p_{z}})=(\frac{\partial}{\partial p_{x}},\frac{\partial}{\partial p_{y}},\frac{\partial}{\partial p_{z}})$. Our units are natural
Heaviside-Lorentz units, $\hbar = c = k_B = \epsilon_0 = \mu_0 =1$. The metric tensor is $g_{\mu \nu}
= \mathrm{diag}(+,-,-,-)$.

\section{DHW functions and their equations of motion \label{sec:DHW-functions}}

In this section we define the DHW functions as expansion
coefficients of the equal-time Wigner function. The choice of the gauge potential
is to some degree arbitrary. Here we use the temporal gauge
$A^{0}=0$, for which the EM fields are given by $\mathbf{E}=-\partial_{t}\mathbf{A}$
and $\mathbf{B}=\boldsymbol{\nabla}\times\mathbf{A}$. In principle
the EM fields include contributions from external fields and contributions
from all charged particles. But in this paper we will focus on the case of an
external field only and neglect the interaction between particles, which
corresponds to a free Fermi gas.

The gauge-invariant Wigner operator is given by
\begin{equation}
\hat{W}(X,P)=\int\frac{d^{4}Y}{(2\pi)^{4}}\exp\bigg(-iy^{\mu}p_{\mu}\bigg)\bar{\psi}\bigg(X+\frac{Y}{2}\bigg)\otimes U\bigg(X+\frac{Y}{2},X-\frac{Y}{2}\bigg)\psi\bigg(X-\frac{Y}{2}\bigg),
\end{equation}
where $\psi$ is the Dirac field operator for spin-1/2 particles.
This formula represents the Fourier transform with respect to the relative
position $Y$ of the direct product of two fermion field operators at space-time points $X+\frac{Y}{2}$
and $X-\frac{Y}{2}$, respectively. The gauge link between these two points renders the Wigner operator
gauge-invariant and is defined as
\begin{equation}
U\bigg(X+\frac{Y}{2},X-\frac{Y}{2}\bigg)=\exp\bigg[-iey^{\mu}
\int_{-1/2}^{1/2}dsA_{\mu}(X+sY)\bigg],
\end{equation}
where $A_{\mu}$ is the gauge potential, e.g.\ in this paper the electromagnetic potential.
Taking the expectation value of the Wigner operator in
a state $\left|\Omega\right\rangle $, we obtain the Wigner function
\begin{equation}
W(X,P)\equiv\langle\Omega|\hat{W}(X,P)|\Omega\rangle.
\label{eq:def-wigner-function}
\end{equation}
The Wigner function, defined in eight-dimensional phase space $(x^{\mu},p^{\mu})$,
is Lorentz covariant but does not have a clear physical interpretation \cite{Vasak:1987um,Elze:1989un}.
By integrating over the energy $p^{0}$ we obtain the corresponding
equal-time Wigner function \cite{BialynickiBirula:1991tx,Zhuang:1998bqx,Gorbar:2017awz},
which can be interpreted as a quasi-probability distribution in six-dimensional
phase space $(\mathbf{x},\mathbf{p})$ at time $t$. Such a procedure
evidently breaks the Lorentz covariance, however, the equation of motion
might simplify to that of an initial-value problem. On the other hand,
we adopt the Hartree approximation, i.e., we treat the quantum EM field
as a semi-classical EM field, from which we derive the following
formula for the equal-time Wigner function $W(t,\mathbf{x},\mathbf{p})$,
\begin{eqnarray}
W(t,\mathbf{x},\mathbf{p}) & = & \int\frac{d^{3}\mathbf{y}}{(2\pi)^{3}}\exp\bigg[i\mathbf{y}\cdot\mathbf{p}
+ie\int_{-1/2}^{1/2}ds\ \mathbf{y}\cdot\mathbf{A}(t,\mathbf{x}+s\mathbf{y})\bigg]
\bigg\langle\Omega\bigg|\bar{\psi}\bigg(t,\mathbf{x}+\frac{\mathbf{y}}{2}\bigg)\otimes
 \psi\bigg(t,\mathbf{x}-\frac{\mathbf{y}}{2}\bigg)\bigg|\Omega\bigg\rangle.
\label{eq:equal-time-form}
\end{eqnarray}
The Hartree approximation ignores higher-loop radiative corrections
and is a good approximation for strong EM fields.

The equation of motion for the equal-time Wigner function can be derived
from the Dirac equation. We consider a non-zero chemical potential $\mu$ associated 
with the conservation of fermion number, which, for the
sake of simplicity, we assume to be constant in space-time, hence its derivatives vanish. An effective way to include
the chemical potential is by adding a term $+ \mu \hat{N}$ to the Dirac--Hamilton operator $\hat{H}_D$,
where $\hat{N}$ is the fermion-number operator. The corresponding Dirac equation reads
\begin{equation}
[i\gamma^{\sigma}(\partial_{\sigma}+ieA_{\sigma})-m+\mu\gamma^{0}]\psi(X)=0.
\end{equation}
Taking the time derivative of Eq.\ (\ref{eq:equal-time-form}) and simplifying the result using the
Dirac equation, we obtain the following equation of motion \cite{Hebenstreit:2010vz}:
\begin{equation}
D_{t}W=\frac{1}{2}\mathbf{D}_{\mathbf{x}}\cdot\left[W,\gamma^{0}\boldsymbol{\gamma}\right]-i
\boldsymbol{\Pi} \cdot \left\{ W,\gamma^{0}\boldsymbol{\gamma}\right\} +im\left[W,\gamma^{0}\right],\label{eq:EOM}
\end{equation}
where the operators $D_{t}$, $\mathbf{D}_{\mathbf{x}}$, and $\boldsymbol{\Pi}$
are generalized operators for time and spatial derivatives, as well as momentum,
in the presence of an EM field,
\begin{eqnarray}
D_{t} & \equiv & \partial_{t}+e\int_{-1/2}^{1/2}ds\ \mathbf{E}(t,\mathbf{x}
-i s \boldsymbol{\nabla}_{\mathbf{p}})\cdot\boldsymbol{\nabla}_{\mathbf{p}},\nonumber \\
\mathbf{D}_{\mathbf{x}} & \equiv & \boldsymbol{\nabla}_{\mathbf{x}}+e\int_{-1/2}^{1/2}ds\ \mathbf{B}(t,\mathbf{x}
-i s \boldsymbol{\nabla}_{\mathbf{p}})\times\boldsymbol{\nabla}_{\mathbf{p}},\nonumber \\
\boldsymbol{\Pi} & \equiv & \mathbf{p}+ie\int_{-1/2}^{1/2}ds\ s\ \mathbf{B}(t,\mathbf{x}
-i s \boldsymbol{\nabla}_{\mathbf{p}})\times\boldsymbol{\nabla}_{\mathbf{p}}.\label{eq:operators}
\end{eqnarray}
For spatially homogeneous EM fields $F^{\mu\nu}(t,\mathbf{x})=F^{\mu\nu}(t)$,
these operators become local,
\begin{eqnarray}
D_{t} & = & \partial_{t}+e\mathbf{E}(t)\cdot\boldsymbol{\nabla}_{\mathbf{p}},\nonumber \\
\mathbf{D}_{\mathbf{x}} & = & \boldsymbol{\nabla}_{\mathbf{x}}+e\mathbf{B}(t)\times\boldsymbol{\nabla}_{\mathbf{p}},
\nonumber \\
\boldsymbol{\Pi} & = & \mathbf{p}.
\label{eq:operators_spathom}
\end{eqnarray}

It can be easily checked that the equal-time Wigner function $W(t,\mathbf{x},\mathbf{p})$
satisfies $W^{\dagger}=\gamma^{0}W\gamma^{0}$ and can be decomposed
in terms of the 16 independent generators of the Clifford algebra
$\Gamma_{i}=\{1,\ i\gamma^{5},\ \gamma^{\mu},\ \gamma^{5}\gamma^{\mu},\ \frac{1}{2}\sigma^{\mu\nu}\}$,
\begin{equation}
W(t,\mathbf{x},\mathbf{p})=\frac{1}{4}\bigg(\mathcal{F}+i\gamma^{5}\mathcal{P}+\gamma^{\mu}\mathcal{V}_{\mu}
+\gamma^{5}\gamma^{\mu}\mathcal{A}_{\mu}+\frac{1}{2}\sigma^{\mu\nu}\mathcal{S}_{\mu\nu}\bigg),
\label{eq:decomposition}
\end{equation}
where $\sigma^{\mu\nu}=\frac{i}{2}\left[\gamma^{\mu},\gamma^{\nu}\right]$
is the anti-symmetric spin tensor. These 16 functions, commonly called Dirac-Heisenberg-Wigner
(DHW) functions, are real functions of time $t$ and six-dimensional
phase space $(\mathbf{x},\mathbf{p})$. The tensor part can be further
decomposed into two vector functions
\begin{equation}
\mathcal{\boldsymbol{T}}=\left( \begin{array}{c} \mathcal{S}^{10} \\
\mathcal{S}^{20}\\ \mathcal{S}^{30}\end{array}\right),\;\;
\mathcal{\boldsymbol{S}}=\left( \begin{array}{c} \mathcal{S}^{23} \\
\mathcal{S}^{31}\\ \mathcal{S}^{12}
\end{array}\right).
\end{equation}
Some of these DHW functions have a clear physical meaning \cite{BialynickiBirula:1991tx},
e.g.\ $\mathcal{F}$ determines the mass density, $\mathcal{V}_{\mu}$
the vector-charge current density, $\mathcal{A}_{\mu}$
the chiral-charge current density, and $\boldsymbol{S}$
the magnetic-moment density. Substituting Eq.\ (\ref{eq:decomposition})
into the equation of motion (\ref{eq:EOM}) and projecting onto the
16 basis matrices, we find a system of partial differential equations (PDEs)
for the DHW functions
\begin{equation}
D_{t}\left(\begin{array}{c}
\boldsymbol{G}_{1}\\
\boldsymbol{G}_{2}\\
\boldsymbol{G}_{3}\\
\boldsymbol{G}_{4}
\end{array}\right)=\left(\begin{array}{cccc}
0 & 0 & 0 & M_{1}\\
0 & 0 & -M_{2} & 0\\
0 & -M_{2} & 0 & -2m\\
-M_{1} & 0 & 2m & 0
\end{array}\right)\left(\begin{array}{c}
\boldsymbol{G}_{1}\\
\boldsymbol{G}_{2}\\
\boldsymbol{G}_{3}\\
\boldsymbol{G}_{4}
\end{array}\right),\label{eq:eom-matrix-form}
\end{equation}
where the DHW functions have been divided into four groups and each
group is composed of four functions \cite{Sheng:2017lfu},
\begin{eqnarray}
\boldsymbol{G}_{1}=\left(\begin{array}{c}
\mathcal{F}\\
\mathcal{\boldsymbol{S}}
\end{array}\right) & , & \boldsymbol{G}_{2}=\left(\begin{array}{c}
\mathcal{V}_{0}\\
\boldsymbol{\mathcal{A}}
\end{array}\right),\nonumber \\
\boldsymbol{G}_{3}=\left(\begin{array}{c}
\mathcal{A}_{0}\\
\boldsymbol{\mathcal{V}}
\end{array}\right) & , & \boldsymbol{G}_{4}=\left(\begin{array}{c}
\mathcal{P}\\
\mathcal{\boldsymbol{T}}
\end{array}\right).\label{eq:def-Gi}
\end{eqnarray}
In Eq.\ (\ref{eq:eom-matrix-form}), we have introduced the two matrices
\begin{eqnarray}
M_{1}\equiv\left(\begin{array}{cc}
0 & 2\boldsymbol{\Pi}^{T}\\
2\boldsymbol{\Pi} & \mathbf{D}_{\mathbf{x}}^{\times}
\end{array}\right) & , & M_{2}\equiv\left(\begin{array}{cc}
0 & \mathbf{D}_{\mathbf{x}}^{T}\\
\mathbf{D}_{\mathbf{x}} & -2\boldsymbol{\Pi}^{\times}
\end{array}\right),\label{eq:matrices}
\end{eqnarray}
where $\boldsymbol{\Pi}$ and $\mathbf{D}_{\mathbf{x}}$ were already defined
in Eq.\ (\ref{eq:operators}). For any three-dimensional column vector
$\mathbf{V}$, $\mathbf{V}^{T}$ is the corresponding transposed
vector (line vector) and $\mathbf{V}^{\times}$ represents the anti-symmetric
$3\times3$ matrix
\begin{equation}
\mathbf{V}^{\times}=\left(\begin{array}{ccc}
0 & -V_{z} & V_{y}\\
V_{z} & 0 & -V_{x}\\
-V_{y} & V_{x} & 0
\end{array}\right),
\end{equation}
the elements of which are $\mathbf{V}^{\times}_{ij}=-\epsilon _{ijk}V_k$.
The differential equations \ref{eq:eom-matrix-form} are equivalent to the ones in 
Refs.~\cite{Zhuang:1998bqx,Hebenstreit:2010vz} but here we write them in a matrix form. 
When dealing with the Landau levels in a constant magnetic field, this matrix form 
allows for more compact formulas \cite{Sheng:2017lfu}.

\section{Spatially homogeneous electric field
\label{sec:In-electric-fields}}

In this section we will simplify the equations of motion (\ref{eq:eom-matrix-form}) for the DHW functions
in a spatially homogeneous electric field and then give the solution for
a constant electric field. The electric field is taken to point into the
$z$-direction. In this case, the gauge potential is $\mathbf{A}(t)=A(t)\mathbf{e}_{z}$
with $\partial_{t}A(t)=-E(t)$.
A similar procedure has been adopted in Ref.\ \cite{Hebenstreit:2010vz}, where 
the authors only discussed the pair production in vacuum. In a thermal environment, the low-energy 
states are occupied, which blocks the production of pairs into these 
states. In this section, a thermal equilibrium 
distribution is assumed at the initial time. Since collisions between particles are not included, all 
existing particles are accelerated by the electric field and thus the distribution depends on the canonical 
momentum. We show that, in the solution, the thermal distribution appears as an overall suppression factor, 
which does not influence the structure of the PDE system. 
The basis used in this section is different from the one in 
Ref.\ \cite{Hebenstreit:2010vz}, but both span the same Hilbert space and thus are 
equivalent to each other. The system of PDEs and corresponding initial conditions derived with the basis in this section provides a convenient framework to describe pair production 
in parallel electric and magnetic fields in Sec.\ \ref{sec:In-parallel-electric}. 

Let us first consider the DHW functions for a free gas of fermions. These can
be derived by first quantizing the field operators in terms of solutions for free
particles, which can be found in any textbook of quantum
field theory, and then inserting the field operators into the definition
of the Wigner function. The result is
\begin{eqnarray}
\left(\begin{array}{c}
\mathcal{F}\\
\boldsymbol{\mathcal{V}}
\end{array}\right)_{\mathrm{free}}(\mathbf{p}) & = & \frac{d_{s}}{(2\pi)^{3}}\frac{1}{E_{\mathbf{p}}}\bigg[
f_{FD}(E_{\mathbf{p}}-\mu)+f_{\mathrm{FD}}(E_{\mathbf{p}}+\mu)-1\bigg] \left(\begin{array}{c} m\\
\mathbf{p}
\end{array}\right),\nonumber \\
\mathcal{V}_{0,\mathrm{free}}(\mathbf{p}) & = & \frac{d_{s}}{(2\pi)^{3}}\bigg[
f_{FD}(E_{\mathbf{p}}-\mu)-f_{\mathrm{FD}}(E_{\mathbf{p}}+\mu)+1\bigg].\label{eq:initial-1}
\end{eqnarray}
here $d_{s}$ is the degeneracy of spin, which is $d_{s}=2$ for spin-$\frac{1}{2}$
particles, and
\begin{equation}
f_{\mathrm{FD}}(E_{\mathbf{p}}\mp\mu)=\frac{1}{1+\exp[\beta(E_{\mathbf{p}}\mp\mu)]}\label{eq:initial-1b}
\end{equation}
is the Fermi-Dirac distribution for particles/anti-particles with energy $E_{\mathbf{p}}$ and vector chemical potential
$\mu$, while $\beta=T^{-1}$ is the
inverse temperature. Note that the fermionic field operators in the definition
(\ref{eq:equal-time-form}) are not normal-ordered, therefore taking the
expectation value in the state $| \Omega \rangle$ yields an additional $\mp 1$, which appears in
the square brackets in Eq.\ (\ref{eq:initial-1}). Here $\mathcal{V}_{0}$,
$\boldsymbol{\mathcal{V}}$, and $\mathcal{F}$ are the charge, current,
and mass densities, respectively. All other DHW functions vanish for a free gas of fermions,
$\mathcal{P} = \mathcal{A}_0 = \boldsymbol{\mathcal{A}} = \mathbf{S} =\mathbf{T} = 0$,
which can be proven using the completeness
relations for the Dirac spinors $u(\mathbf{k},s)$ and $v(\mathbf{k},s)$.

We now proceed to solve the equations of motion (\ref{eq:eom-matrix-form}) for the
DHW functions.
Due to the absence of a magnetic field and translation invariance of
the system, we can set the spatial derivative $\mathbf{D}_{\mathbf{x}}$ to zero and $\boldsymbol{\Pi} \equiv \mathbf{p}$.
The matrices in Eq.\ (\ref{eq:matrices}) then simplify to
\begin{eqnarray}
M_{1}=\left(\begin{array}{cc}
0 & 2\mathbf{p}^{T}\\
2\mathbf{p} & \mathbf{0}_{3\times3}
\end{array}\right) & , & M_{2}=\left(\begin{array}{cc}
0 & \mathbf{0}_{1\times3}\\
\mathbf{0}_{3\times1} & -2\mathbf{p}^{\times}
\end{array}\right),\label{eq:matrices-1}
\end{eqnarray}
and $D_{t}=\partial_{t}+eE(t)\partial_{p_{z}}$. Then, the
16 equations of motion for the DHW functions can be divided into several groups.
The equation for the charge density separates from the others and reads
\begin{equation}
D_{t}\mathcal{V}_{0}(t,\mathbf{p})=0.
\end{equation}
After integrating over the momentum $\mathbf{p}$ and neglecting the boundary
terms (because there is no particle with infinite $p_z$), the above equation is nothing but the conservation
of net charge. Furthermore, the ten equations of motion for the DHW functions $\mathcal{F},\boldsymbol{\mathcal{V}},
\boldsymbol{\mathcal{A}},$ and $\mathcal{\boldsymbol{T}}$
decouple from the other five for the functions $\mathcal{P}, \mathcal{A}_0,$ and $\mathbf{S}$.
These latter ones will no longer be considered, because their initial values are zero and thus
they will remain zero for later times as well. In matrix form we have
\begin{equation}
D_{t}\boldsymbol{w}(t,\mathbf{p})=M(\mathbf{p})\boldsymbol{w}(t,\mathbf{p}),\label{eq:Dt-w}
\end{equation}
where $\boldsymbol{w}(t,\mathbf{p})=\left(\mathcal{F},\,\boldsymbol{\mathcal{V}},\, 
\boldsymbol{\mathcal{A}},\,\mathcal{\boldsymbol{T}}\right)^T$ is a ten-dimensional vector 
consisting of ten DHW functions
and $M(\mathbf{p})$ is a $10\times10$ matrix
\begin{equation}
M(\mathbf{p})=2\left(\begin{array}{cccc}
0 & 0 & 0 & \mathbf{p}^{T}\\
0 & 0 & \mathbf{p}^{\times} & -m\\
0 & \mathbf{p}^{\times} & 0 & 0\\
-\mathbf{p} & m & 0 & 0
\end{array}\right).
\end{equation}
Inspired by the form (\ref{eq:initial-1}) of the free DHW functions, we make the following ansatz for
the solution of Eq.\ (\ref{eq:Dt-w}),
\begin{eqnarray}
\boldsymbol{w}(t,\mathbf{p}) & = & \frac{d_{s}}{(2\pi)^{3}}\bigg\{
f_{FD}\left( E_{\mathbf{p}+e\delta A(t)\mathbf{e}_{z}}- \mu \right)
+{f}_{\mathrm{FD}}\left(E_{\mathbf{p}+e\delta A(t)\mathbf{e}_{z}}+ \mu\right)-1\bigg\} \sum_{i=1}^{3}\chi_{i}(t,\mathbf{p})
\boldsymbol{e}_{i}(\mathbf{p}_{T}).\label{eq:expand}
\end{eqnarray}
Here $\delta A(t)\equiv A(t)-A(t_{0})$ is the difference of the gauge
potentials at time $t$ and at initial time $t_{0}$.
The distribution thus depends on the canonical momentum, which reflects the acceleration of fermions in 
an electric field. Since acting the operator $D_t$ on $\mathbf{p}+e\delta A(t)\mathbf{e}_{z}$ gives zero,
 the term in the curly brackets in Eq.\ (\ref{eq:expand}) behaves like a constant overall factor and can be taken out of Eq.\ (\ref{eq:Dt-w}). The value of this term is in the range $(-1,0)$, 
which is the effect of Pauli blocking by particles already present in the thermal system. 
Note that, since the matrix $\boldsymbol{w}(t,\mathbf{p})$
has dimension ten, in principle we would need ten basis vectors $\boldsymbol{e}_{i}$ in the ansatz (\ref{eq:expand}).
However, we actually only need three because these form a closed
sub-space under the operators $D_{t}$ and $M(\mathbf{p})$, while
the initial conditions are also inside this sub-space. These basis vectors are
\begin{equation}
\boldsymbol{e}_{1}=\left(\begin{array}{c}
0\\
\mathbf{e}_{z}\\
\mathbf{0}\\
\mathbf{0}
\end{array}\right), \;\; \boldsymbol{e}_{2}(\mathbf{p}_{T}) = \frac{1}{m_{T}}\left(\begin{array}{c}
m\\
\mathbf{p}_{T}\\
\mathbf{0}\\
\mathbf{0}
\end{array}\right), \;\; \boldsymbol{e}_{3}(\mathbf{p}_{T})=\frac{1}{m_{T}}\left(\begin{array}{c}
0\\
\mathbf{0}\\
\mathbf{e}_{z}\times\mathbf{p}_{T}\\
-m\mathbf{e}_{z}
\end{array}\right),
\end{equation}
which are independent of $t$ and $p_{z}$, so that $D_{t}\boldsymbol{e}_{i}=0$
for all $i=1,2,3$. Here we have introduced the transverse mass $m_{T}\equiv\sqrt{m^{2}+\mathbf{p}_{T}^{2}}$,
so that the three basis vectors are properly normalized, $\boldsymbol{e}_{i}\cdot\boldsymbol{e}_{j}=\delta_{ij}$.
We can also check that they are closed under the operator $M(\mathbf{p})$,
\begin{equation}
M(\mathbf{p})\left(\begin{array}{c}
\boldsymbol{e}_{1}\\
\boldsymbol{e}_{2}\\
\boldsymbol{e}_{3}
\end{array}\right)=2\left(\begin{array}{ccc}
0 & 0 & -m_{T}\\
0 & 0 & p_{z}\\
m_{T} & -p_{z} & 0
\end{array}\right)\left(\begin{array}{c}
\boldsymbol{e}_{1}\\
\boldsymbol{e}_{2}\\
\boldsymbol{e}_{3}
\end{array}\right).\label{eq:Me-Dt-e-1}
\end{equation}
Inserting the ansatz (\ref{eq:expand}) into Eq.\ (\ref{eq:Dt-w})
and using Eq.\ (\ref{eq:Me-Dt-e-1}) we can derive the equations of motion for
the coefficient functions $\chi_i(t)$, %
\begin{eqnarray}
D_{t} \left( \begin{array}{c} \chi_{1} \\ \chi_2 \\ \chi_3 \end{array} \right) (t,\mathbf{p})
& = & 2 \left( \begin{array}{ccc} 0 & 0 & m_T \\
0 & 0 & - p_z \\
- m_T & p_z & 0 \end{array}
\right)  \left( \begin{array}{c} \chi_{1} \\ \chi_2 \\ \chi_3 \end{array} \right) (t,\mathbf{p}) .\label{eq:PDE}
\end{eqnarray}
In order to solve this system of PDEs, we need to specify the initial condition.
Here we choose the values of the DHW functions in the absence of an electric
field. For an integrable electric field, which vanishes sufficiently rapidly for $t \rightarrow \pm \infty$,
such as the Sauter-type field $E(t)=E_{0}\, \mathrm{cosh}^{-2}(t/\tau)$,
we specify the initial condition for $t_{0}\rightarrow-\infty$, where we take the DHW functions to assume the
values given by Eq.\ (\ref{eq:initial-1}).

However, for a constant electric field $E(t)=E_{0}$, the momentum
shift will be infinitely large if we take $t_{0}\rightarrow-\infty$,
because a constant field is not integrable. In reality, fermions
will collide with each other, kinetic energy will be converted
to thermal energy, and the system will approach thermodynamical equilibrium.
Here we make the assumption that the system
is already in thermodynamical equilibrium at
initial time $t_{0}$. We should find a solution that coincides with Eq.\
(\ref{eq:initial-1}) when the field strength is sufficiently small, $E_{0}\rightarrow0$, i.e.,
\begin{equation}
\left.\left(\begin{array}{c}
\chi_{1}\\
\chi_{2}\\
\chi_{3}
\end{array}\right)(t,\mathbf{p})\right|_{E_{0}\rightarrow0}=\frac{1}{E_{\mathbf{p}}}\left(\begin{array}{c}
p_{z}\\
m_{T}\\
0
\end{array}\right).\label{eq:initial-2}
\end{equation}

The pair-production rate and the corresponding Wigner function have
analytical solutions for both a constant field and a Sauter-type
field, see Ref.\ \cite{Hebenstreit:2010vz} for details of the derivation
from quantum kinetic theory, which we will not repeat here. In a constant
field $E(t)=E_{0}$ the solutions do not depend on space-time coordinates,
which is obvious because of translation invariance,
\begin{equation}
\left(\begin{array}{c}
\chi_{1}\\
\chi_{2}\\
\chi_{3}
\end{array}\right)(\mathbf{p})=\left(\begin{array}{c}
d_{1}\bigg(\eta,\sqrt{\frac{2}{eE_{0}}}p_{z}\bigg)\\
\frac{m_{T}}{\sqrt{2eE_{0}}}\, d_{2}\bigg(\eta,\sqrt{\frac{2}{eE_{0}}}p_{z}\bigg)\\
\frac{m_{T}}{\sqrt{2eE_{0}}}\, d_{3}\bigg(\eta,\sqrt{\frac{2}{eE_{0}}}p_{z}\bigg)
\end{array}\right),\label{eq:solutions}
\end{equation}
where $\eta\equiv m_{T}^{2}/(eE_{0})$ is the dimensionless transverse
mass square and the auxiliary functions are listed in Eq.\ (\ref{eq:dfunctions})
of App.\ \ref{sec:Auxiliary-functions}. It is easy to check numerically that
the solutions (\ref{eq:solutions}) satisfy the constraint
(\ref{eq:initial-2}) and the system (\ref{eq:PDE}) of PDEs. The
corresponding DHW functions can be readily derived by inserting Eq.\
(\ref{eq:solutions}) into Eq.\ (\ref{eq:expand}).

\section{Parallel and spatially homogeneous electric and magnetic fields\label{sec:In-parallel-electric}}

In this section, we will consider spatially homogeneous electric and magnetic fields which are parallel
to each other. Without loss of generality, the fields are assumed to point into the $z$-direction.
We also assume the magnetic field to be constant in time. Then the solution can be simplified by separately
considering the different Landau levels. We provide an analytical solution
for the case when the electric field is also constant in time.

\subsection{Initial conditions}

Analogously to the case without magnetic field, we choose the DHW
functions in a pure magnetic field as initial condition
for the system (\ref{eq:eom-matrix-form}) of PDEs. Since we consider this field to be constant in space and time,
an analytical solution can be found.
The corresponding covariant DHW functions in eight-dimensional
phase space have been determined in Ref.\ \cite{Sheng:2017lfu}. In this paper,
we set the axial chemical potential to zero, i.e., we do not consider the chiral magnetic effect.
Then, using the results of Ref.\ \cite{Sheng:2017lfu} the covariant DHW functions read
\begin{eqnarray}
\left(\begin{array}{c}
\boldsymbol{G}_{1}(P)\\
\boldsymbol{G}_{2}(P)
\end{array}\right) & = & \sum_{n=0}V^{(n)}(p_{0},p_{z})\boldsymbol{e}_{1}^{(n)}(p_{T})\left(\begin{array}{c}
m\\
p_{0}+\mu
\end{array}\right),\nonumber \\
\boldsymbol{G}_{3}(P) & = & p_{z}V^{(0)}(p_{0},p_{z})\boldsymbol{e}_{1}^{(0)}(p_{T})
+\sum_{n>0}V^{(n)}(p_{0},p_{z})\bigg[p_{z}\boldsymbol{e}_{2}^{(n)}(p_{T})
+\sqrt{2neB}\boldsymbol{e}_{3}^{(n)}(\mathbf{p}_{T})\bigg],\nonumber \\
\boldsymbol{G}_{4}(P) & = & 0,\label{eq:sol-Wigner-function-1}
\end{eqnarray}
where
\begin{equation}
V^{(n)}(p_{0},p_{z})=\frac{2(2-\delta_{n0})}{(2\pi)^{3}}\delta\bigg\{(p_{0}+\mu)^{2}-[E_{p_{z}}^{(n)}]^{2}\bigg\}
\bigg\{\theta(p_{0}+\mu)f_{FD}(p_{0})+\theta(-p_{0}-\mu)\bigg[f_{FD}(-p_{0})-1\bigg]\bigg\}.
\end{equation}
Here, $E_{p_{z}}^{(n)}=\sqrt{m^{2}+p_{z}^{2}+2neB}$ is the energy
of the $n$th Landau level in a constant magnetic field and $f_{FD}$ is the
Fermi-Dirac distribution function. The basis vectors $\boldsymbol{e}_{i}^{(n)}$
are given in Eq.\ (\ref{eq:ein}) of the appendix.
Since the pair production is a dynamical process, it is more convenient to use the equal-time formula. 
We emphasize that the covariant DHW functions can be obtained from the equal-time ones by applying an additional 
Fourier transformation in $t$, i.e., $t\rightarrow p_0$, and conversely, the equal-time DHW functions can 
be derived from the covariant ones by integrating over $p_{0}$. Here we give the equal-time DHW functions,
\begin{eqnarray}
\boldsymbol{G}_{1}(\mathbf{p}) & = & \sum_{n=0}\frac{m}{E_{p_{z}}^{(n)}}C_{1}^{(n)}(p_{z})\boldsymbol{e}_{1}^{(n)}(p_{T}),
\nonumber \\
\boldsymbol{G}_{2}(\mathbf{p}) & = & \sum_{n=0}C_{2}^{(n)}(p_{z})\boldsymbol{e}_{1}^{(n)}(p_{T}),\nonumber \\
\boldsymbol{G}_{3}(\mathbf{p}) & = & \frac{p_{z}}{E_{p_{z}}^{(0)}}C_{1}^{(0)}(p_{z})\boldsymbol{e}_{1}^{(0)}(p_{T})
+\sum_{n>0}C_{1}^{(n)}(p_{z})\frac{1}{E_{p_{z}}^{(n)}}\bigg[p_{z}\boldsymbol{e}_{2}^{(n)}(p_{T})
+\sqrt{2neB}\boldsymbol{e}_{3}^{(n)}(\mathbf{p}_{T})\bigg],\nonumber \\
\boldsymbol{G}_{4}(\mathbf{p}) & = & 0.\label{eq:initial condtions}
\end{eqnarray}
Here, $C_{1}^{(n)}(p_{z})\equiv\int dp_{0}E_{p_{z}}^{(n)}V^{(n)}(p_{0},p_{z})$
and $C_{2}^{(n)}(p_{z})\equiv\int dp_{0}(p_{0}+\mu)V^{(n)}(p_{0},p_{z})$, respectively. The $p_0$-integrals
can be performed, yielding the result
\begin{eqnarray}
C_{1}^{(n)}(p_{z}) & = & \frac{2-\delta_{n0}}{(2\pi)^{3}}\bigg[f_{FD}(E_{p_{z}}^{(n)}-\mu)
+f_{FD}(E_{p_{z}}^{(n)}+\mu)-1\bigg],\nonumber \\
C_{2}^{(n)}(p_{z}) & = & \frac{2-\delta_{n0}}{(2\pi)^{3}}\bigg[f_{FD}(E_{p_{z}}^{(n)}-\mu)
-f_{FD}(E_{p_{z}}^{(n)}+\mu)+1\bigg].\label{eq:C1-C2}
\end{eqnarray}
The Fermi-Dirac distributions are the number densities in coordinate space
for fermions/anti-fermions. The prefactor $2-\delta_{n0}$ is the
spin degeneracy of the various Landau levels.
Comparing with Eq.\ (\ref{eq:initial-1}) without the electromagnetic field, 
Eq.\ (\ref{eq:initial condtions}) has more non-vanishing components and 
depends on the Landau levels $n$. We will show later that in 
a constant magnetic field, different Landau levels evolve independently.

\subsection{Equations of Motion}

In the presence of a constant magnetic field, the operator for the generalized spatial
differentiation, cf.\ second Eq.\ (\ref{eq:operators_spathom}), becomes
$\mathbf{D}_{\mathbf{x}}=e\mathbf{B} \times \boldsymbol{\nabla}_{\mathbf{p}}$,
where the ordinary spatial gradient $\boldsymbol{\nabla}_{\mathbf{x}}$ has been dropped, since all
considered fields are spatially homogeneous and the system is translation-invariant.

The lowest Landau level is special since we only
need the basis vector $\boldsymbol{e}_{1}^{(0)}(p_T)$ to describe the dynamics in the 
lowest Landau level.
The reason is that, in a constant magnetic field,
$\boldsymbol{e}_{1}^{(0)}(p_T)$ is an eigenvector for all
operators $D_{t}$, $M_{1}$, $M_{2}$ appearing in the equation of motion (\ref{eq:eom-matrix-form}),
\begin{equation}
M_{1}\boldsymbol{e}_{1}^{(0)}(p_{T})=2p_{z}\boldsymbol{e}_{1}^{(0)}(p_{T}),\ M_{2}\boldsymbol{e}_{1}^{(0)}(p_{T})
=D_{t}\boldsymbol{e}_{1}^{(0)}(p_{T})=0.\label{eq:M_e_Dt_e}
\end{equation}

For the higher Landau levels, the situation is more complicated.
In the last subsection we have shown that the basis vectors $\boldsymbol{e}_{i}^{(n)}$,
$i=1,2,3$, cf.\ Eq.\ (\ref{eq:ein}), are necessary to describe the equal-time DHW
functions in a constant magnetic field. One can easily check that these
basis vectors are not closed under the operator $M_{2}$ defined in Eq.\
(\ref{eq:matrices}). In order to construct a closed space under $M_{2}$,
we need another basis vector, $\boldsymbol{e}_{4}^{(n)}$,
the definition of which is also given in Eq.\ (\ref{eq:ein}). Acting with $M_{1},M_{2}$
onto these basis vectors and using the relations (\ref{eq:differential-Lambda-1})
gives for all higher Landau levels $n>0$
\begin{eqnarray}
M_{1}\boldsymbol{e}_{i}^{(n)}(\mathbf{p}_{T}) & = & \sum_{j=1}^4 (c_{1}^{(n)})_{ij}^{T}
\boldsymbol{e}_{j}^{(n)}(\mathbf{p}_{T}),
\nonumber \\
M_{2}\boldsymbol{e}_{i}^{(n)}(\mathbf{p}_{T}) & = & \sum_{j=1}^4(c_{2}^{(n)})_{ij}^{T}
\boldsymbol{e}_{j}^{(n)}(\mathbf{p}_{T}),
\label{eq:M_e}
\end{eqnarray}
where the coefficient matrices are
\begin{eqnarray}
c_{1}^{(n)}\equiv2\left(\begin{array}{cccc}
0 & p_{z} & \sqrt{2neB} & 0\\
p_{z} & 0 & 0 & 0\\
\sqrt{2neB} & 0 & 0 & 0\\
0 & 0 & 0 & 0
\end{array}\right) & , & c_{2}^{(n)}\equiv-2\left(\begin{array}{cccc}
0 & 0 & 0 & 0\\
0 & 0 & 0 & \sqrt{2neB}\\
0 & 0 & 0 & -p_{z}\\
0 & -\sqrt{2neB} & p_{z} & 0
\end{array}\right). \label{eq:c-i}
\end{eqnarray}
Note that the transpose of these matrices appears in Eq.\ (\ref{eq:M_e}).

We have already seen in Eq.\ (\ref{eq:initial condtions}) that, when the electric field vanishes,
the DHW functions can be expressed
in terms of the basis vectors $\boldsymbol{e}_{i}^{(n)}$.
Taking Eq.\ (\ref{eq:initial condtions})
as initial condition one can straightforwardly conclude that the DHW
functions will stay in the space spanned by $\boldsymbol{e}_{i}^{(n)}$ when they evolve according to
the equation of motion (\ref{eq:eom-matrix-form}). This is because $D_t$ acting on $\boldsymbol{e}_{i}^{(n)}$
gives zero, while we have already seen that these basis vectors form a closed subset when acting with
$M_{1,2}$ onto them, see Eq.\ (\ref{eq:M_e}).
We thus make the ansatz
\begin{equation}
\boldsymbol{G}_{i}(t,\mathbf{p})=f_{i}^{(0)}(t,p_{z})\boldsymbol{e}_{1}^{(0)}(p_{T})+\sum_{n>0}
\sum_{j=1}^{4}f_{ij}^{(n)}(t,p_{z})\boldsymbol{e}_{j}^{(n)}(\mathbf{p}_{T}),\label{eq:G-i}
\end{equation}
where $i,j=1,2,3,4$.
Since the magnetic field is assumed to be constant in time,
the basis vectors $\boldsymbol{e}_{i}^{(n)}$ are also independent of time. Inserting
Eq.\ (\ref{eq:G-i}) into the equation of motion (\ref{eq:eom-matrix-form}) for the DHW functions,
and using the orthogonality relations (\ref{eq:orth-enem-1}) and (\ref{eq:orth-e0e0-ene0}) for
the basis vectors,
we can derive the equations of motion for the functions $f_{i}^{(0)}$ and $f_{ij}^{(n)}$. For the lowest Landau level
we obtain

\begin{equation}
D_{t}\left(\begin{array}{c}
f_{1}^{(0)}\\
f_{2}^{(0)}\\
f_{3}^{(0)}\\
f_{4}^{(0)}
\end{array}\right)(t,p_{z})=2\left(\begin{array}{cccc}
0 & 0 & 0 & p_{z}\\
0 & 0 & 0 & 0\\
0 & 0 & 0 & -m\\
-p_{z} & 0 & m & 0
\end{array}\right)\left(\begin{array}{c}
f_{1}^{(0)}\\
f_{2}^{(0)}\\
f_{3}^{(0)}\\
f_{4}^{(0)}
\end{array}\right)(t,p_{z}).\label{eq:LLL-1}
\end{equation}
The equations for the higher levels are
\begin{equation}
D_{t}\left(\begin{array}{c}
\boldsymbol{f}_{1}^{(n)}\\
\boldsymbol{f}_{2}^{(n)}\\
\boldsymbol{f}_{3}^{(n)}\\
\boldsymbol{f}_{4}^{(n)}
\end{array}\right)(t,p_{z})=\left(\begin{array}{cccc}
0 & 0 & 0 & c_{1}^{(n)}\\
0 & 0 & -c_{2}^{(n)} & 0\\
0 & -c_{2}^{(n)} & 0 & -2m\\
-c_{1}^{(n)} & 0 & 2m & 0
\end{array}\right)\left(\begin{array}{c}
\boldsymbol{f}_{1}^{(n)}\\
\boldsymbol{f}_{2}^{(n)}\\
\boldsymbol{f}_{3}^{(n)}\\
\boldsymbol{f}_{4}^{(n)}
\end{array}\right)(t,p_{z}),\label{eq:HLL}
\end{equation}
where $\boldsymbol{f}_{i}^{(n)}\equiv(f_{i1}^{(n)},f_{i2}^{(n)},f_{i3}^{(n)},f_{i4}^{(n)})^T$
is a four-dimensional column vector. We observe that, on account of the
orthogonality relations (\ref{eq:orth-enem-1}), (\ref{eq:orth-e0e0-ene0}), the equations
for the different Landau levels separate from each other, which greatly facilitates
the solution of the equations of motion.

\subsection{Lowest Landau level}

The spin of the fermion in the lowest Landau level with positive/negative charge
is parallel/anti-parallel to the magnetic field. The equation for $f_{2}^{(0)}$,
cf.\  the second line in Eq.\ (\ref{eq:LLL-1}),
decouples from the other equations and gives rise to the conservation of
net fermion number in the lowest Landau level. In order to see this, we note that the net fermion number
density $\mathcal{V}_{0}(t,\mathbf{p})$ is the first component of
$\boldsymbol{G}_{2}$ in Eq.\ (\ref{eq:def-Gi}). The lowest Landau
level contributes just $f_{2}^{(0)}(t,p_{z})\Lambda_{+}^{(0)}(p_{T})$, cf.\ Eqs.\ (\ref{eq:G-i}) and (\ref{eq:ein}).
Acting with $D_{t}=\partial_{t}+eE_{0}\partial_{p_{z}}$ on that and integrating
over $\mathbf{p}$ yields with the definition $n^{(0)} \equiv \int d^{3}\mathbf{p} \, f_{2}^{(0)}(t,p_{z})
\Lambda_{+}^{(0)}(p_{T})$ the conservation law
\begin{equation}
\partial_{t}n^{(0)}=\int d^{3}\mathbf{p}\bigg[D_{t}f_{2}^{(0)}(t,p_{z})\bigg]\Lambda_{+}^{(0)}(p_{T})=0,
\end{equation}
where we have integrated by parts and neglected
the boundary term. The equation $D_{t}f_{2}^{(0)}(t,p_{z})=0$,
together with $\left.f_{2}^{(0)}(t,p_{z})\right|_{E_{0}\rightarrow0}=C_{2}^{(n)}(p_{z})$
has the special solution
\begin{equation}
f_{2}^{(0)}(t,p_{z})=C_{2}^{(0)}(p_{z}-eE_{0}t).
\end{equation}
This solution describes an overall acceleration of all charged particles.
We note that in this paper we focus on a free fermion gas, so there
are no collisions to prevent the acceleration.

The equations of motion for the other three functions $f_{i=1,3,4}^{(0)}$ are coupled with each
other. In order to simplify the problem, we make an ansatz which splits off the thermal
distribution functions,
\begin{equation}
\left\{ f_{1}^{(0)},f_{3}^{(0)},f_{4}^{(0)}\right\} =\left\{ \chi_{1}^{(0)},\chi_{2}^{(0)},\chi_{3}^{(0)}\right\}
C_{1}^{(0)}(p_{z}-eE_{0}t).\label{eq:sep-thermal}
\end{equation}
where $C_{1}^{(0)}$ is defined in Eq.\ (\ref{eq:C1-C2}).
Here, $p_{z}+eA(t) = p_z - eE_0 t$ is the canonical momentum. When acting with $D_{t}$
on $f_{i}^{(0)}$, we only need to consider its effect on $\chi_{i}$, because $D_{t}C_{1}^{(0)}(p_{z}-eE_0t)=0$.
Thus, the equations of motion for the $\chi_{i}$ are the same as the one for the corresponding
$f_i^{(0)}$, cf.\ Eq.\ (\ref{eq:LLL-1}),
\begin{equation}
D_{t}\left(\begin{array}{c}
\chi_{1}^{(0)}\\
\chi_{2}^{(0)}\\
\chi_{3}^{(0)}
\end{array}\right)(t,p_{z})=2\left(\begin{array}{ccc}
0 & 0 & p_{z}\\
0 & 0 & -m\\
-p_{z} & m & 0
\end{array}\right)\left(\begin{array}{c}
\chi_{1}^{(0)}\\
\chi_{2}^{(0)}\\
\chi_{3}^{(0)}
\end{array}\right)(t,p_{z}).\label{eq:LLL-chi1}
\end{equation}
Comparing the ansatz (\ref{eq:sep-thermal}) with the initial
condition (\ref{eq:initial condtions}), i.e., for $E_0 \rightarrow 0$, we find
\begin{equation}
\left.\left(\begin{array}{c}
\chi_{1}^{(0)}\\
\chi_{2}^{(0)}\\
\chi_{3}^{(0)}
\end{array}\right)(t,p_{z})\right|_{E_{0}\rightarrow0}=\frac{1}{E_{p_{z}}^{(0)}}\left(\begin{array}{c}
m\\
p_{z}\\
0
\end{array}\right),\label{eq:initial}
\end{equation}

The system (\ref{eq:LLL-chi1}) of PDEs with the initial condition
(\ref{eq:initial}) coincides with the PDE system (\ref{eq:PDE}) in a pure electric
field (substituting $\chi_1 \rightarrow \chi_2^{(0)}$, $\chi_2 \rightarrow \chi_1^{(0)}$, $\chi_3 \rightarrow -\chi_3^{(0)}$
and setting $p_T^{2}=0$).
One can therefore immediately give the solution
for a constant electric field $E(t)=E_{0}$,
\begin{equation}
\left(\begin{array}{c}
\chi_{1}^{(0)}\\
\chi_{2}^{(0)}\\
\chi_{3}^{(0)}
\end{array}\right)(p_{z})=\left(\begin{array}{c}
\frac{m}{\sqrt{2eE_{0}}}d_{2}\bigg(\eta^{(0)},\sqrt{\frac{2}{eE_{0}}}p_{z}\bigg)\\
d_{1}\bigg(\eta^{(0)},\sqrt{\frac{2}{eE_{0}}}p_{z}\bigg)\\
-\frac{m}{\sqrt{2eE_{0}}}d_{3}\bigg(\eta^{(0)},\sqrt{\frac{2}{eE_{0}}}p_{z}\bigg)
\end{array}\right), \label{eq:chi-123}
\end{equation}
with $d_{i}$ defined in Eq.\ (\ref{eq:dfunctions}) and $\eta^{(0)}=m^{2}/eE_{0}$.
Multiplying Eq.\ (\ref{eq:chi-123}) with $C_{1}^{(0)}\left(p_{z}-eE_0t)\right)$ gives
the functions $ f_{1}^{(0)},f_{3}^{(0)},$ and $f_{4}^{(0)}$
in a constant electric field,
\begin{equation}
\left(\begin{array}{c}
f_{1}^{(0)}\\
f_{3}^{(0)}\\
f_{4}^{(0)}
\end{array}\right)(p_{z})=\left(\begin{array}{c}
\frac{m}{\sqrt{2eE_{0}}}d_{2}\bigg(\eta^{(0)},\sqrt{\frac{2}{eE_{0}}}p_{z}\bigg)C_{1}^{(0)}(p_{z}-eE_{0}t)\\
d_{1}\bigg(\eta^{(0)},\sqrt{\frac{2}{eE_{0}}}p_{z}\bigg)C_{1}^{(0)}(p_{z}-eE_{0}t)\\
-\frac{m}{\sqrt{2eE_{0}}}d_{3}\bigg(\eta^{(0)},\sqrt{\frac{2}{eE_{0}}}p_{z}\bigg)C_{1}^{(0)}(p_{z}-eE_{0}t)
\end{array}\right).\label{eq:LLL-solution}
\end{equation}
Inserting these functions into Eq.\ (\ref{eq:G-i}), one obtains
the contribution from the lowest Landau level to the DHW functions.

\subsection{Higher Landau levels}

For the higher Landau levels ($n>0$) we can read off from Eqs.\ (\ref{eq:initial condtions}), (\ref{eq:G-i})
that, when switching off the electric
field, the only functions which do not vanish are $f_{11}^{(n)}$, $f_{21}^{(n)}$, $f_{32}^{(n)}$, and $f_{33}^{(n)}$.
Writing down the equations of motion (\ref{eq:HLL}) for the $f_{ij}^{(n)}$ functions for the higher Landau levels
using Eq.\ (\ref{eq:c-i}) we observe that $f_{24}^{(n)}$, $f_{42}^{(n)}$, and $f_{43}^{(n)}$ couple with
$f_{11}^{(n)}$, $f_{32}^{(n)}$, and $f_{33}^{(n)}$ in the presence
of an electric field. The corresponding six basis functions form a
closed sub-space. The other nine functions are decoupled to three independent groups,
$\left\{f_{12}^{(n)},f_{13}^{(n)}, f_{31}^{(n)}, f_{41}^{(n)}\right\}$,
$\left\{f_{22}^{(n)}, f_{23}^{(n)}, f_{34}^{(n)}, f_{44}^{(n)}\right\}$ and
$\left\{f_{14}^{(n)}\right\}$,
each forms a closed set of homogeneous PDEs. However, since all
of them have vanishing values when the electric field is zero, all of them will stay zero during the further
evolution, even after switching on the electric field.

In the following, we therefore focus on the seven
non-trivial functions $f_{11}^{(n)}$, $f_{21}^{(n)}$, $f_{24}^{(n)}$, $f_{32}^{(n)}$,
$f_{33}^{(n)}$, $f_{42}^{(n)}$, and $f_{43}^{(n)}$.
The equation for $f_{21}^{(n)}$, $D_{t}f_{21}^{(n)}=0$, decouples from the others. As
discussed in the previous subsection,
this equation is nothing but the conservation of net charge in each Landau
level. The solution is
\begin{equation}
f_{21}^{(n)}(t,p_{z})=C_{2}^{(n)}(p_{z}-eE_{0}t),\label{eq:charge}
\end{equation}
where $p_z - eE_0t$ describes the overall acceleration of all existing
particles by the electric field in the $z$-direction.

As already mentioned above, the other six functions, $f_{11}^{(n)}$, $f_{24}^{(n)}$, $f_{32}^{(n)}$, $f_{33}^{(n)}$,
$f_{42}^{(n)}$, $f_{43}^{(n)}$,
satisfy a six-dimensional system of PDEs. They can be further decoupled by
introducing the following linear combinations,
\begin{equation}
\left(\begin{array}{cc}
g_{1}^{(n)} & g_{3}^{(n)}\\
g_{4}^{(n)} & g_{2}^{(n)}
\end{array}\right)=\frac{1}{m^{(n)}}\left(\begin{array}{cc}
m & \sqrt{2neB}\\
\sqrt{2neB} & -m
\end{array}\right)\left(\begin{array}{cc}
f_{11}^{(n)} & f_{24}^{(n)}\\
f_{33}^{(n)} & f_{42}^{(n)}
\end{array}\right),\label{eq:transformation}
\end{equation}
where the effective mass at level $n$ is $m^{(n)}\equiv\sqrt{m^{2}+2neB}$.
Then we get the following two groups of equations
\begin{equation}
D_{t}\left(\begin{array}{c}
g_{1}^{(n)}\\
g_{2}^{(n)}\\
f_{32}^{(n)}
\end{array}\right)(t,p_{z})=2\left(\begin{array}{ccc}
0 & -p_{z} & 0\\
p_{z} & 0 & -m^{(n)}\\
0 & m^{(n)} & 0
\end{array}\right)\left(\begin{array}{c}
g_{1}^{(n)}\\
g_{2}^{(n)}\\
f_{32}^{(n)}
\end{array}\right)(t,p_{z}),
\end{equation}
and
\begin{equation}
D_{t}\left(\begin{array}{c}
g_{3}^{(n)}\\
g_{4}^{(n)}\\
f_{43}^{(n)}
\end{array}\right)(t,p_{z})=2\left(\begin{array}{ccc}
0 & -p_{z} & 0\\
p_{z} & 0 & m^{(n)}\\
0 & -m^{(n)} & 0
\end{array}\right)\left(\begin{array}{c}
g_{3}^{(n)}\\
g_{4}^{(n)}\\
f_{43}^{(n)}
\end{array}\right)(t,p_{z}), \label{eq:2nd-group}
\end{equation}
In this way, $g_{1}^{(n)}$, $g_{2}^{(n)}$, and $f_{32}^{(n)}$ decouple from
$g_{3}^{(n)}$, $g_{4}^{(n)}$, and $f_{43}^{(n)}$.
When the electric field
vanishes, we find from Eqs.\ (\ref{eq:initial condtions}), (\ref{eq:G-i}), and (\ref{eq:transformation}) that
$g_{3}^{(n)}$, $g_{4}^{(n)}$, and $f_{43}^{(n)}$ vanish. Under the time
evolution determined by Eq.\ (\ref{eq:2nd-group}) this will remain the case after switching on $\mathbf{E}$.
Therefore, we only need to focus on the equations for $g_{1}^{(n)}$, $g_{2}^{(n)}$, and $f_{32}^{(n)}$.
Analogous to the treatment of the lowest Landau level, we assume that the
solutions have the following form,
\begin{equation}
\left\{ g_{1}^{(n)},g_{2}^{(n)},f_{32}^{(n)}\right\} =\left\{ \chi_{1}^{(n)},\chi_{2}^{(n)},\chi_{3}^{(n)}\right\}
C_{1}^{(n)}(p_{z}-eE_{0}t).\label{eq:assume}
\end{equation}
Since $D_t C_1^{(n)}(p_{z}-eE_{0}t) =0$, the system of PDEs for $\{\chi_{1}^{(n)},\chi_{2}^{(n)},\chi_{3}^{(n)}\}$
reads
\begin{equation}
D_{t}\left(\begin{array}{c}
\chi_{1}^{(n)}\\
\chi_{2}^{(n)}\\
\chi_{3}^{(n)}
\end{array}\right)(t,p_{z})=2\left(\begin{array}{ccc}
0 & -p_{z} & 0\\
p_{z} & 0 & -m^{(n)}\\
0 & m^{(n)} & 0
\end{array}\right)\left(\begin{array}{c}
\chi_{1}^{(n)}\\
\chi_{2}^{(n)}\\
\chi_{3}^{(n)}
\end{array}\right)(t,p_{z}).\label{eq:HLL-1}
\end{equation}
The initial values can be deduced by first reading off the
functions $f_{ij}^{(n)}$ via a comparison of Eq.\ (\ref{eq:initial condtions})
with Eq.\ (\ref{eq:G-i}) and then using Eq.\ (\ref{eq:transformation}),
\begin{equation}
\left.\left(\begin{array}{c}
\chi_{1}^{(n)}\\
\chi_{2}^{(n)}\\
\chi_{3}^{(n)}
\end{array}\right)(t,p_{z})\right|_{E_{0}\rightarrow0}=\frac{1}{E_{p_{z}}^{(n)}}\left(\begin{array}{c}
m^{(n)}\\
0\\
p_{z}
\end{array}\right).\label{eq:initial-3}
\end{equation}
The system (\ref{eq:HLL-1}) of PDEs and the initial condition (\ref{eq:initial-3})
coincide with the PDE system (\ref{eq:PDE}) in a pure electric field (replacing
$\chi_1 \rightarrow \chi_3^{(n)}$, $\chi_2 \rightarrow \chi_1^{(n)}$, $\chi_3 \rightarrow \chi_2^{(n)}$,
and setting $p_T^2 = 2neB$). Then the solutions for a constant
electric field $E(t)=E_{0}$ are straightforward to write down,
\begin{equation}
\left(\begin{array}{c}
\chi_{1}^{(n)}\\
\chi_{2}^{(n)}\\
\chi_{3}^{(n)}
\end{array}\right)(\mathbf{p})=\left(\begin{array}{c}
\frac{m^{(n)}}{\sqrt{2eE_{0}}}d_{2}\bigg(\eta^{(n)},\sqrt{\frac{2}{eE_{0}}}p_{z}\bigg)\\
\frac{m^{(n)}}{\sqrt{2eE_{0}}}d_{3}\bigg(\eta^{(n)},\sqrt{\frac{2}{eE_{0}}}p_{z}\bigg)\\
d_{1}\bigg(\eta^{(n)},\sqrt{\frac{2}{eE_{0}}}p_{z}\bigg)
\end{array}\right),\label{eq:solutions-1-1}
\end{equation}
with $d_{i}$ defined in Eq.\ (\ref{eq:dfunctions}) and $\eta^{(n)}=(m^{2}+2neB)/(eE_{0})$.

Now that we have found the solution for the $\chi_{i}^{(n)}$, we can insert it into the ansatz
(\ref{eq:assume}) and obtain $g_{1}^{(n)}$, $g_{2}^{(n)}$, and $f_{32}^{(n)}$.
Then using the inverse of the transformation (\ref{eq:transformation}),
one can compute all non-vanishing functions,
\begin{eqnarray}
\left(\begin{array}{c}
f_{11}^{(n)}\\
f_{33}^{(n)}
\end{array}\right) & = & \left(\begin{array}{c}
m\\
\sqrt{2neB}
\end{array}\right)\frac{1}{\sqrt{2eE_{0}}}\, d_{2}\bigg(\eta^{(n)},\sqrt{\frac{2}{eE_{0}}}p_{z}\bigg)C_{1}^{(n)}(p_{z}-eE_{0}t),
\nonumber \\
\left(\begin{array}{c}
f_{24}^{(n)}\\
f_{42}^{(n)}
\end{array}\right) & = & \left(\begin{array}{c}
\sqrt{2neB}\\
-m
\end{array}\right)\frac{1}{\sqrt{2eE_{0}}}\, d_{3}\bigg(\eta^{(n)},\sqrt{\frac{2}{eE_{0}}}p_{z}\bigg)C_{1}^{(n)}(p_{z}-eE_{0}t),
\nonumber \\
f_{32}^{(n)} & = & d_{1}\bigg(\eta^{(n)},\sqrt{\frac{2}{eE_{0}}}p_{z}\bigg)C_{1}^{(n)}(p_{z}-eE_{0}t),\label{eq:HLL-solution}
\end{eqnarray}
together with $f_{21}^{(n)}$ from Eq.\ (\ref{eq:charge}).
The remaining ten functions are zero.

\section{Pair-production rate}

\label{sec:Pair-production-rate}

In the last section we have derived the DHW functions in constant
electric and magnetic fields. In this section we will relate the DHW
functions to pair production. Note that, in the presence of an electric field,
the system cannot remain in thermodynamical equilibrium.

Let us first consider a multi-pair system, where the particles
are described by the plane-wave solutions of the free Dirac equation.
Inserting these wave functions into the definition of the Wigner function
and then projecting onto the unit matrix and the gamma matrices $\boldsymbol{\gamma}$,
we obtain the contribution of fermion/anti-fermion pairs to the DHW
functions \cite{BialynickiBirula:1991tx},
\begin{eqnarray} \label{eq:F-V}
\mathcal{F}=\frac{2m}{E_{\mathbf{p}}}\left[n_{\mathrm{pair}}(\mathbf{p}) -1 \right],
&  &\boldsymbol{\mathcal{V}}=\frac{2\mathbf{p}}{E_{\mathbf{p}}}\left[n_{\mathrm{pair}}(\mathbf{p}) -1 \right],
\end{eqnarray}
where $n_{\mathrm{pair}}(\mathbf{p})$ is the number density of pairs in phase space.
The Pauli principle implies that $0 \leq n_{\mathrm{pair}}(\mathbf{p}) \leq 1$.
The density of pairs will change due to the pair-production process 
caused by the electric field. The corresponding rate is given by
\begin{equation} 
\label{eq:dn_pair-dt}
\frac{d}{dt}n_{\mathrm{pair}}=\frac{1}{2}\frac{d}{dt}\int d^{3}\mathbf{p}\frac{m\mathcal{F}
+\mathbf{p}\cdot\boldsymbol{\mathcal{V}}}{E_{\mathbf{p}}},
\end{equation}
where $n_{\mathrm{pair}} = \int d^3\mathbf{p}\, n_{\mathrm{pair}}(\mathbf{p})$ is the number of pairs.
Equation (\ref{eq:dn_pair-dt}) can be proven by inserting Eq.\ (\ref{eq:F-V}) into the right-hand side.

Analogously, for a multi-pair system in a constant background magnetic field, the on-shell energy is
$E_{p_{z}}^{(n)}=\sqrt{m^{2}+p_{z}^{2}+2neB}$. If there is pair production by an electric field in the system, its rate in the $n$th Landau level can then be calculated via
\begin{equation}
\frac{d}{dt}n_{\mathrm{pair}}^{(n)}=\frac{1}{2}\frac{d}{dt}\int d^{3}\mathbf{p}\frac{m\mathcal{F}^{(n)}
+\mathbf{p}\cdot\boldsymbol{\mathcal{V}}^{(n)}}{E_{p_{z}}^{(n)}}.
\end{equation}
Here, $\mathcal{F}^{(n)}$ and $\boldsymbol{\mathcal{V}}^{(n)}$ represent
the DHW functions corresponding to the $n$th Landau level. Employing Eq.\ (\ref{eq:def-Gi})
and the ansatz (\ref{eq:G-i}), we get
\begin{equation}
\frac{d}{dt}n_{\mathrm{pair}}^{(n)}=\frac{1}{2}\frac{d}{dt}\int d^{3}\mathbf{p}\bigg[\frac{\eta^{(n)}}{E_{p_{z}}^{(n)}}
\sqrt{\frac{eE_{0}}{2}}d_{2}\bigg(\eta^{(n)},\sqrt{\frac{2}{eE_{0}}}p_{z}\bigg)+\frac{p_{z}}{E_{p_{z}}^{(n)}}
d_{1}\bigg(\eta^{(n)},\sqrt{\frac{2}{eE_{0}}}p_{z}\bigg)\bigg]C_{1}^{(n)}(p_{z}-eE_{0}t)\Lambda_{+}^{(n)}(p_{T}),
\end{equation}
where $C_{1}^{(n)}$ is given by Eq.\ (\ref{eq:C1-C2}). The integration
over $\mathbf{p}_{T}$ can be performed using Eq.\ (\ref{eq:int-Lambda}).
Replacing the kinetic momentum $p_{z}$ by the canonical momentum
$q_{z}=p_{z}-eE_{0}t$ we obtain the pair-production rate in the $n$th
Landau level in parallel electric and magnetic fields and a thermal background,
\begin{equation}
\frac{d}{dt}n_{\mathrm{pair}}^{(n)}=\int dq_{z}\bigg[1-f_{\mathrm{FD}}(E_{q_{z}}^{(n)}-\mu)-f_{\mathrm{FD}}(E_{q_{z}}^{(n)}
+\mu)\bigg]\frac{d}{dt}n_{\mathrm{vac}}^{(n)}(t,q_{z}),
\label{eq:pair-production-rate}
\end{equation}
where $\frac{d}{dt}n_{\mathrm{vac}}^{(n)}(p_{z})$ is the pair-production
rate in vacuum for given quantum numbers $p_{z}$ and $n$,
\begin{eqnarray}
\frac{d}{dt}n_{\mathrm{vac}}^{(n)}(t,q_{z}) & = & -\left(1-\frac{\delta_{n0}}{2}\right)\frac{e^{2}BE_{0}}{(2\pi)^{2}}
\frac{d}{dq_{z}}\left\{ \frac{\eta^{(n)}}{E_{q_{z}+eE_{0}t}^{(n)}}\sqrt{\frac{eE_{0}}{2}}d_{2}
\bigg[\eta^{(n)},\sqrt{\frac{2}{eE_{0}}}(q_{z}+eE_{0}t)\bigg]\right.\nonumber \\
 &  & \hspace*{3.5cm} \left.+\frac{q_{z}+eE_{0}t}{E_{q_{z}+eE_{0}t}^{(n)}}
 d_{1}\bigg[\eta^{(n)},\sqrt{\frac{2}{eE_{0}}}(q_{z}+eE_{0}t)\bigg] \right\} .
\end{eqnarray}
The Fermi-Dirac distributions in the square bracket in Eq.\ (\ref{eq:pair-production-rate})
describe the suppression of pair production due to the Pauli exclusion
principle. Summing Eq.\ (\ref{eq:pair-production-rate}) over all
Landau levels yields the total pair-production rate.

In a medium where the chemical potential is zero but the temperature
is nonzero the suppression factor is $1-2f_{\mathrm{FD}}(E_{q_{z}}^{(n)})=\tanh\frac{\beta E_{q_{z}}^{(n)}}{2}$,
which suppresses the production of pairs with small energies. This
factor agrees with the result of Ref.\ \cite{Kim:2008em}. However,
the integral in Eq.\ (\ref{eq:pair-production-rate})
is hard to calculate numerically, because the auxiliary functions $d_{1}$ and $d_{2}$
are highly oscillatory at large $q_{z}$, which makes the integration
converge too slowly. This problem can be solved by separating the
vacuum contribution from the thermal contribution. The pair-production
rate in the $n$th Landau level in vacuum can be analytically calculated, using the asymptotic behavior
of the $d_{1,2}$ functions, cf.\ App.\ \ref{sec:Auxiliary-functions},
\begin{eqnarray}
\frac{d}{dt}n_{\mathrm{vac}}^{(n)}=\int dq_{z}\frac{d}{dt}n_{\mathrm{vac}}^{(n)}(t,q_{z})
& = & -\left(1-\frac{\delta_{n0}}{2}\right)\frac{e^{2}BE_{0}}{(2\pi)^{2}}\bigg[d_{1}\bigg(\eta^{(n)},-\infty\bigg)
+d_{1}\bigg(\eta^{(n)},+\infty\bigg)\bigg]\nonumber \\
 & = & \left(1-\frac{\delta_{n0}}{2}\right)\frac{e^{2}BE_{0}}{2\pi^{2}}\exp\bigg(-\pi\frac{m^{2}+2neB}{eE_{0}}\bigg).
\end{eqnarray}
The total rate from all Landau levels is
\begin{equation}
\frac{d}{dt}\sum_{n=0}^{\infty}n_{\mathrm{vac}}^{(n)}
=\frac{e^{2}E_{0}B}{4\pi^{2}}\exp\left(-\frac{\pi m^{2}}{eE_{0}}\right)
\coth\left(\frac{\pi B}{E_{0}}\right),
\end{equation}
which was previously derived in Refs.\ \cite{Nikishov:1969tt,Bunkin:1970iz,Dunne:2004nc}.
We see that the rate will be enhanced for $B\gg E_0$
compared to that without the magnetic field \cite{Schwinger:1951nm}.
Similarly we can also derive the production rate of chiral charge $dn_5/dt$
in a strong magnetic field, which gives the anomaly with
the pair production \cite{Fukushima:2010vw,Copinger:2018ftr}.

\begin{figure}
\includegraphics[scale=0.4]{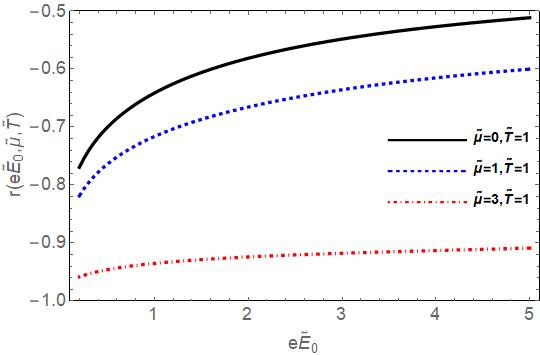}
\hspace{0.2cm}
\includegraphics[scale=0.4]{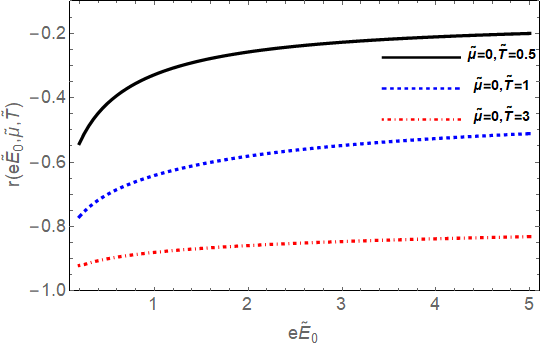}

\caption{The ratio of the thermal contribution to the total pair-production
rate to the vacuum contribution for a constant electric field. Left panel:
dependence on electric field strength for temperature $\tilde{T}=1$
and chemical potential $\tilde{\mu}=0,\ 1,\ 3$. Right panel:
$\tilde{\mu}=0$ and $\tilde{T}=0.5,\ 1,\ 3$. }
\label{Fig1}
\end{figure}

The thermal contribution in Eq. (\ref{eq:pair-production-rate})
for the $n$th Landau level is
\begin{equation}
\frac{d}{dt}n_{\mathrm{thermal}}^{(n)}=-\int dq_{z}\bigg[f_{\mathrm{FD}}(E_{q_{z}}^{(n)}-\mu)
+f_{\mathrm{FD}}(E_{q_{z}}^{(n)}+\mu)\bigg]\frac{d}{dt}n_{\mathrm{vac}}^{(n)}(t,q_{z}).
\end{equation}
The Fermi--Dirac distributions provide an exponential suppression $\sim e^{-E_{q_{z}}^{(n)}}$ for
large $q_z$, thus the $q_{z}$-integral converges quickly. In order to
show the thermal suppression in a physically intuitive way, we introduce the ratio
$r$ of the thermal to the vacuum contribution.
This ratio is a function of time $t$ and the three dimensionless
parameters $e\tilde{E}_{0}\equiv\frac{eE_{0}}{[m^{(n)}]^{2}}$, $\tilde{T}\equiv\frac{T}{m^{(n)}}$,
and $\tilde{\mu}\equiv\frac{\mu}{m^{(n)}}$, where $m^{(n)}=\sqrt{m^{2}+2neB}$
is the effective mass in the $n$th Landau level. The total pair-production
rate in the $n$th Landau level is given by
\begin{equation}
\frac{d}{dt}n_{\mathrm{pair}}^{(n)}=\bigg[1+r(t,\tilde{E}_{0},\tilde{T},\tilde{\mu})\bigg]\frac{d}{dt}n_{\mathrm{vac}}^{(n)}.
\end{equation}
In order to show the thermal influence on pair production, we choose
the time $t=0$, which is when the canonical momentum
equals the kinetic one. Figure \ref{Fig1} shows the function $r(0,\tilde{E}_{0},\tilde{T},\tilde{\mu})$
at finite dimensionless temperature and chemical potential. The values
stay between $-1$ and $0$ for all parameters considered, which
describes the thermal suppression of pair production as demanded by the
Pauli exclusion principle: a quantum state has a higher probability
to be occupied at higher temperature or higher chemical potential;
this occupation will block the production of new pairs with the same
quantum numbers. When the electric field is strong enough, pairs with
higher energies, which have smaller thermal occupation numbers, are more likely
to be excited. Thus the suppression is inversely proportional to the
electric field strength.

\section{Summary\label{sec:Summary}}

In this paper we have analytically calculated the Wigner function as well
as the Schwinger pair production in constant and parallel electric and magnetic
fields. We have derived the equation of motion for the equal-time Wigner function,
whose sixteen components, the so-called DHW functions, have definite physical meanings.
One can relate the Schwinger pair-production rate to some of these functions.
For the case of a pure
constant electric field, we took the vacuum
values for the sixteen DHW functions as initial condition. Then, we obtained an
analytic solution for the
system of PDEs for the DHW functions. For
parallel electric and magnetic fields, we adopted a similar method to
calculate the DHW functions. We showed that the
contributions of different Landau levels separate from each other. Under the replacement
$p_{T}^{2}\rightarrow 2neB$
the system of PDEs and the condition when the electric field vanishes
coincide with those in a pure electric field for each Landau level. This provides us
with a new method for calculating
the pair production in parallel electric and magnetic fields. Analytical
solutions for the DHW functions for the case of constant electric and magnetic fields, together
with the pair-production rate in each Landau level are derived. Our results
can be directly generalised to the case of finite temperature and chemical
potential. The calculation shows that the pair-production rate is
thermally suppressed and the suppression is proportional to the thermodynamic
variables $T$ and $\mu$. More energetic pairs can be created in a 
stronger electric field, which are less likely to be Pauli-blocked by the
thermal distribution, and this leads to a decrease of the suppression factor.

The equation of motion for the Wigner function is equivalent to the
Dirac equation if we adopt the classical-field approximation. However,
the Wigner function contains sixteen independent components, which leads
to a sixteen-dimensional system of PDEs. Due to advances in computer
technology in the past few decades, it becomes possible to numerically solve
this PDE system in some simplified cases. In this paper, we have
found a set of basis functions in the presence of a constant magnetic field.
These basis functions provide us with a way to replace the continuous transverse
momenta $p_{x}$, $p_{y}$ by the discrete Landau level index
$n$. The parameter space of the Wigner function is then simplified from
six-dimensional phase space $(\mathbf{x},\mathbf{p})$ to the four-dimensional space
spanned by $(\mathbf{x},p_{z})$ plus one discrete parameter $n$, which
makes the system of PDEs more amenable for a numerical solution. However,
the case considered in this paper, i.e., homogeneous and parallel electromagnetic
fields, is effectively only a $(1+1)$-dimensional problem, whereas the fields
in real experiments are more likely to be space-time dependent. Nevertheless,
the way of decomposing the Wigner function presented here may inspire future works
and may be a convenient starting point for the Wigner-function approach.

\textit{Acknowledgments.}
QW is supported in part by the 973 program under Grant No. 2015CB856902
and by NSFC under Grant No. 11535012.
XLS  is supported in part by China Scholarship Council and the Deutsche Forschungsgemeinschaft (DFG)
through the grant CRC-TR 211 "Strong-interaction matter under extreme conditions".
DHR acknowledges support by the High-End Visiting Expert project
of the State Administration of Foreign Experts Affairs (SAFEA) of China
and by the Deutsche Forschungsgemeinschaft (DFG, German Research Foundation)
through the Collaborative Research Center CRC-TR 211 ``Strong-interaction matter
under extreme conditions'' -- project number 315477589 - TRR 211.

\appendix

\section{Auxiliary functions\label{sec:Auxiliary-functions}}

\begin{figure}
\includegraphics[scale=0.4]{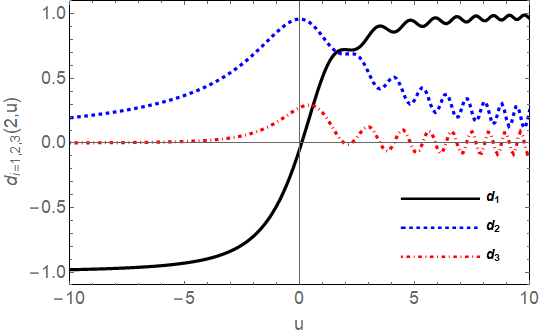}
\includegraphics[scale=0.4]{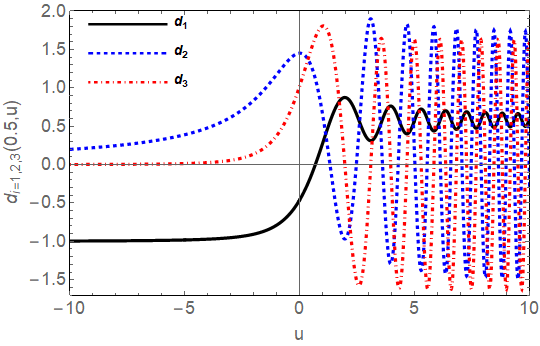}

\caption{$u$-dependence of the auxiliary functions $d_{i}(\eta,u)$, $i=1,\ 2,\ 3$
for $\eta=2$ (left panel) and $\eta = 0.5$ (right panel).}
\label{Fig2}
\end{figure}

We have introduced three auxiliary functions in the solutions for the DHW functions,
\begin{eqnarray}
d_{1}(\eta,u) & = & -1+e^{-\frac{\pi\eta}{4}}\eta\bigg|D_{-1-i\eta/2}(-ue^{i\frac{\pi}{4}})\bigg|^{2},\nonumber \\
d_{2}(\eta,u) & = & e^{-\frac{\pi\eta}{4}}e^{i\frac{\pi}{4}}D_{-1-i\eta/2}(-ue^{i\frac{\pi}{4}})
D_{i\eta/2}(-ue^{-i\frac{\pi}{4}})+c.c.,\nonumber \\
d_{3}(\eta,u) & = & e^{-\frac{\pi\eta}{4}}e^{-i\frac{\pi}{4}}D_{-1-i\eta/2}(-ue^{i\frac{\pi}{4}})
D_{i\eta/2}(-ue^{-i\frac{\pi}{4}})+c.c.,\label{eq:dfunctions}
\end{eqnarray}
where $D_{\nu}$ is the parabolic cylinder function. These functions satisfy the following differential equations,
\begin{eqnarray}
\frac{d}{du}d_{1}(\eta,u) & = & \eta d_{3}(\eta,u),\nonumber \\
\frac{d}{du}d_{2}(\eta,u) & = & -ud_{3}(\eta,u),\nonumber \\
\frac{d}{du}d_{3}(\eta,u) & = & -2d_{1}(\eta,u)+ud_{2}(\eta,u).
\end{eqnarray}
We plot the $d_{i}$ as
function of $u$ in Fig.\ \ref{Fig2}. We observe that
that all these functions are convergent for $u\rightarrow-\infty$,
but only $d_{1}$ is obviously convergent for $u\rightarrow+\infty$.
The functions $d_{2}$ and $d_{3}$ are highly oscillatory in a
finite region for large $u$, thus $d_{2}/u$ and $d_{3}/u$ converge
to zero when $u\rightarrow+\infty$.  Moreover, we have
\begin{eqnarray}
\lim_{u\rightarrow-\infty}d_{1}(\eta,u)=-1 & , & \lim_{u\rightarrow+\infty}d_{1}(\eta,u)=1-2e^{-\pi\eta}.
\end{eqnarray}

Four groups of basis vectors are used in the expansion of the DHW functions in a
constant magnetic field. They are functions of the Landau-level index $n$ and the
transverse momentum $\mathbf{p}_{T}$, or its modulus $p_{T}=\sqrt{\mathbf{p}_{T}\cdot\mathbf{p}_{T}}$,
respectively,
\begin{eqnarray}
\boldsymbol{e}_{1}^{(n)}(p_{T})=\left(\begin{array}{c}
\Lambda_{+}^{(n)}(p_{T})\\
\mathbf{0}_{T}\\
\Lambda_{-}^{(n)}(p_{T})
\end{array}\right), &  & \boldsymbol{e}_{2}^{(n)}(p_{T})=\left(\begin{array}{c}
\Lambda_{-}^{(n)}(p_{T})\\
\mathbf{0}_{T}\\
\Lambda_{+}^{(n)}(p_{T})
\end{array}\right),\nonumber \\
\boldsymbol{e}_{3}^{(n)}(\mathbf{p}_{T})=\frac{\sqrt{2neB}}{p_{T}^{2}}\Lambda_{+}^{(n)}(p_{T})\left(\begin{array}{c}
0\\
\mathbf{p}_{T}\\
0
\end{array}\right), &  & \boldsymbol{e}_{4}^{(n)}(\mathbf{p}_{T})=\frac{\sqrt{2neB}}{p_{T}^{2}}
\Lambda_{+}^{(n)}(p_{T})\left(\begin{array}{c}
0\\
-p_{y}\\
p_{x}\\
0
\end{array}\right).\label{eq:ein}
\end{eqnarray}
Here, the $\Lambda_{\pm}^{(n)}$ functions are defined as
\begin{equation}
\Lambda_{\pm}^{(n)}(p_{T})\equiv\begin{cases}
(-1)^{n}\bigg[L_{n}\bigg(\frac{2p_{T}^{2}}{eB}\bigg)\mp L_{n-1}\bigg(\frac{2p_{T}^{2}}{eB}\bigg)\bigg]
\exp\bigg(-\frac{p_{T}^{2}}{eB}\bigg), & n>0,\\
2\exp\bigg(-\frac{p_{T}^{2}}{eB}\bigg), & n=0,
\end{cases}\label{eq:lambda-n}
\end{equation}
where $L_{n}(x)$ is the $n$th Laguerre polynomial. For the lowest
Landau level, $n=0$, we have $\boldsymbol{e}_{3}^{(0)}=\boldsymbol{e}_{4}^{(0)}=0$
and $\boldsymbol{e}_{1}^{(0)}=\boldsymbol{e}_{2}^{(0)}$. These basis vectors
allow us to separate the $\mathbf{p}_{T}$ dependence. When integrating
over transverse momentum $\mathbf{p}_{T}$, $\Lambda_{+}^{(n)}(p_{T})$
gives the density of states for Landau level $n$, while $\Lambda_{-}^{(n)}(p_{T})$
gives zero for all $n>0$,
\begin{eqnarray}
\frac{1}{(2\pi)^{2}}\int d^{2}\mathbf{p}_{T}\Lambda_{+}^{(n)}(p_{T}) & = & \frac{eB}{2\pi}\nonumber \\
\frac{1}{(2\pi)^{2}}\int d^{2}\mathbf{p}_{T}\Lambda_{-}^{(n)}(p_{T}) & = & 0,\ \ (n\neq0).\label{eq:int-Lambda}
\end{eqnarray}
The basis vectors $\boldsymbol{e}_{i}^{(n)}$ for $i=1,2,3,4$ and $n=0,1,2,\cdots$
are orthogonal with respect to an inner product,
\begin{equation}
\int d^{2}\mathbf{p}_{T}\boldsymbol{e}_{i}^{(m)T}(\mathbf{p}_{T})\boldsymbol{e}_{j}^{(n)}(\mathbf{p}_{T})
=2\pi eB\delta_{mn}\delta_{ij},\label{eq:orth-enem-1}
\end{equation}
for $n>0$, together with
\begin{eqnarray}
\int d^{2}\mathbf{p}_{T}\boldsymbol{e}_{1}^{(0)T}(p_{T})\boldsymbol{e}_{1}^{(0)}(p_{T}) & = & 4\pi eB,\nonumber \\
\int d^{2}\mathbf{p}_{T}\boldsymbol{e}_{i}^{(n)T}(\mathbf{p}_{T})\boldsymbol{e}_{1}^{(0)}(p_{T}) & = & 0.
\label{eq:orth-e0e0-ene0}
\end{eqnarray}
We can also check that the functions $\Lambda_{\pm}^{(n)}$ in Eq.\
(\ref{eq:lambda-n}) satisfy the following relations,
\begin{eqnarray}
eB\partial_{p_{x}}\Lambda_{+}^{(n)}(p_{T}) & = & -2p_{x}\Lambda_{-}^{(n)}(p_{T}),\nonumber \\
eB\partial_{p_{x}}\Lambda_{-}^{(n)}(p_{T}) & = & -2p_{x}\bigg(1-\frac{2neB}{p_{T}^{2}}\bigg)\Lambda_{+}^{(n)}(p_{T}),
\label{eq:differential-Lambda-1}
\end{eqnarray}
which are used to derive Eqs.\ (\ref{eq:M_e_Dt_e}) and (\ref{eq:M_e}).

\bibliographystyle{apsrev}
\bibliography{ref}

\end{document}